\documentstyle[12pt,epsf]{article}
\newcommand{\la}[1]{\label{#1}}

\newcommand{\be}{\begin{equation}}
\newcommand{\ee}{\end{equation}}
\newcommand{\ba}{\begin{eqnarray}}
\newcommand{\ea}{\end{eqnarray}}
\newcommand{\rmi}[1]{{\mbox{\scriptsize #1}}}

\newcommand{\nr}[1]{(\ref{#1})}

\newcommand{\roots}{\sqrt{s}}

\newcommand{\gev}{{\rm GeV}}
\newcommand{\psat}{p_{\rmi{sat}}}
\newcommand{\pqcd}{{\rm pQCD}}

\def\lsim{\raise0.3ex\hbox{$<$\kern-0.75em\raise-1.1ex\hbox{$\sim$}}}
\def\gsim{\raise0.3ex\hbox{$>$\kern-0.75em\raise-1.1ex\hbox{$\sim$}}}

\begin{document}

\begin{titlepage}
\begin{flushright}
HIP-2002-10/TH\\
NORDITA-2002-20-HE\\
hep-ph/0204034\\
3 April, 2002\\
\end{flushright}
\begin{centering}
\vfill

{\bf

        RAPIDITY DEPENDENCE OF PARTICLE PRODUCTION IN ULTRARELATIVISTIC
                     NUCLEAR COLLISIONS
}

\vspace{0.5cm}
 K.J. Eskola$^{\rm a,c,}$\footnote{kari.eskola@phys.jyu.fi},
 K. Kajantie$^{\rm b,c,}$\footnote{keijo.kajantie@helsinki.fi},
P.V. Ruuskanen$^{\rm a,c,}$\footnote{vesa.ruuskanen@phys.jyu.fi} and
K. Tuominen$^{\rm d,}$\footnote{kimmo.tuominen@nordita.dk}

\vspace{1cm}
{\em $^{\rm a}$ Department of Physics, University of Jyv\"askyl\"a, \\P.O. Box 35, FIN-40351
Jyv\"askyl\"a, Finland\\}
\vspace{0.3cm}
{\em $^{\rm b}$ Department of Physics,\\
P.O. Box 64, FIN-00014 University of Helsinki, Finland\\}
\vspace{0.3cm}
{\em $^{\rm c}$ Helsinki Institute of Physics,\\
P.O. Box 64, FIN-00014 University of Helsinki, Finland\\}
\vspace{0.3cm}
{\em $^{\rm d}$ Nordita, \\Blegdamsvej 17, 2100
Copenhagen {\O }, Denmark\\}

\vspace{1cm}
{\bf Abstract}

\end{centering}

\vspace{0.3cm}\noindent We compute the rapidity dependence of particle
and transverse energy production in ultrarelativistic heavy ion
collisions at various beam energies and atomic numbers using the
perturbative QCD + saturation model. The distribution is a broad
gaussian near $y=0$ but the rapid increase of particle production with
the beam energy will via energy conservation strongly constrain the
rapidity distribution at large $y$.

\vfill

\end{titlepage}

\section{Introduction}
The Relativistic Heavy Ion Collider RHIC has already produced a lot of
data at $\roots$ up to 200 GeV and the planning for the ALICE/Large
Hadron Collider experiment at $\roots$ up to 5500 GeV is under way.
The first data are on various large cross section phenomena like
charged multiplicity at zero (pseudo)rapidity in nearly central
collisions \cite{phobos1,phobos3} or at varying \cite{phenix,phobos2}
impact parameter, or as a function of both rapidity and centrality
\cite{phobos4,brahms}.  Much theoretical effort, reviewed in
\cite{eskola}, has been devoted both to predict the multiplicities
before measurements and to draw conclusions from the completed
measurements. With this information the reliability of predictions for
the LHC energies, $\roots=5500$ GeV, is considerably enhanced.

One of the models, reasonably successful in predicting the data at
RHIC energies, is the saturation+pQCD model \cite{ekrt}, based on
microscopic 2$\to$2 partonic processes.  It is actually a member of a
large class of models in which there is one dominant transverse
momentum scale, $\psat$, determined by different versions of a
saturation condition \cite{glr,mq,bm,gm,guo}.  The purpose of this
note is to apply the model to a study of the rapidity dependence of
particle and transverse energy production. Related work in \cite{kl}
and \cite{kln}, based on microscopic 2$\to$1 processes, will be
compared with.

The main observation is that, not surprisingly, energy conservation
places strong constraints \cite{werner} to the domain of validity of
models based on independent 2$\to$2 subcollisions. There is no problem
at zero rapidity: $N(y=0)$ and $E_T(y=0)$ grow rapidly with $\roots$,
but still much more slowly that the total available energy
$\roots$. At larger rapidities such a rapid growth of $N(y)$ cannot be
sustained, the total energy carried by the produced particles would
surpass the available total energy indicating that one has entered a
new domain in which the independent scattering model no more is
valid. We estimate that at RHIC the saturation+pQCD model could be
valid up to $y=1...2$, and at LHC up to $y=3...4$. Within this range
the rapidity distribution is very flat, practically a wide gaussian
with calculable curvature near $y=0$. This is in marked contrast to
the $2\to1$ model in \cite{kl,kln}, where $N(y)$ is sharply peaked at
$y=0$, in fact, has a discontinuous derivative there.
\vspace{0.5cm}

\section{Rapidity dependence in the saturation + pQCD model}
We shall first carry out an approximate analytic discussion
emphasizing parametric dependences.
One starts from the perturbatively determined 
leading-order two-jet cross
section (in standard notation):
\begin{equation}
\frac{d\sigma^{AA\rightarrow kl+X}_\pqcd}{dp_T^2dy_1dy_2}
= K\sum_{ij}\, x_1f_{i/A}(x_1,p_T^2
)\, x_2f_{j/A}(x_2,p_T^2)\,
\sum_{kl}\frac{d\hat\sigma}{d\hat t}^{ij\rightarrow kl}.
\label{minijets}
\end{equation}
To maintain the framework of \cite{ekrt}, we use the GRV94
leading-order parton distribution functions \cite{GRV,PDFLIB}
combined with the nuclear effects (shadowing) from the EKS98
parametrization \cite{EKS98}. An effective factor $K\approx2$ is to
simulate the NLO contributions which are discussed in more detail in
\cite{ET00}.  After integration over $y_2$ Eq.~(\ref{minijets}) gives a
single-jet cross section behaving near the dominant saturation scale
(which depends on $\sqrt s$ and $A$, and also on $y$)
parametrically like \cite{ekt2} \be {d\sigma_\pqcd\over
dyd^2p_T}\sim\roots^{\,2\delta} {\alpha_s^2(p_T^2)\over p_T^4}
e^{-y^2/2\sigma^2(p_T)}, \la{singlejet} \ee where $\delta\approx0.5$
(when no shadowing is included)
and the $y$-dependence has been approximated by a gaussian (the letter
$\sigma$ is used for both the cross section and the width of the
gaussian).  A further integration over $p_T$ from some lower limit
gives the hard cross section (see Figs.~\ref{viuhka} and \ref{satu}
below) \be
\frac{d\sigma_\pqcd}{dy}(p_0)\sim \roots^{\,2\delta}
{\alpha_s^2(p_0^2)\over p_0^2}
e^{-y^2/2\sigma^2(p_0)}.
\ee

Consider now central A+A collisions with the nuclear overlap function
$T_{AA}(b=0)=A^2/(\pi R_A^2)$. The saturation condition then can be
formulated as
\ba
\frac{dN}{dy}(p_0)&=&{A^2\over\pi R_A^2}
\frac{d\sigma_\pqcd}{dy}(p_0)=p_0^2R_A^2
\\ \la{satcond}
&\sim&{A^2\over\pi R_A^2} \roots^{\,2\delta}
{\alpha_s^2(p_0^2)\over p_0^2}
e^{-y^2/2\sigma^2(p_0)},\nonumber
\ea
the solution of which gives $p_0=\psat(A,\sqrt s,y)$. The saturation
is here defined as the geometric saturation on the transverse plane of
the two final state particles in the 2$\to$2 collision.  The $2\to1$
models describe saturation as the separate saturation of the
distribution functions of the two initial state particles. As long as
there is one dominant tranverse momentum scale, the models lead to
similar results, at least when final state particle production in
nearly central collisions is considered.

The principle underlying the saturation condition is illustrated by
Fig.~\ref{mantra}. The perturbative cross sections, of course, are
valid only for large $p_T$. However, one is only interested in the
integral over $p_T$, not the value of the distribution at small
$p_T$. The value of the full integral can also be reproduced by
integrating the perturbative distribution from a lower limit given by
the saturation condition.

\begin{figure}[htb]
\vspace{5cm}
\hspace{0.5cm}
\epsfysize=12cm
\centerline{\hspace{2cm}\epsffile{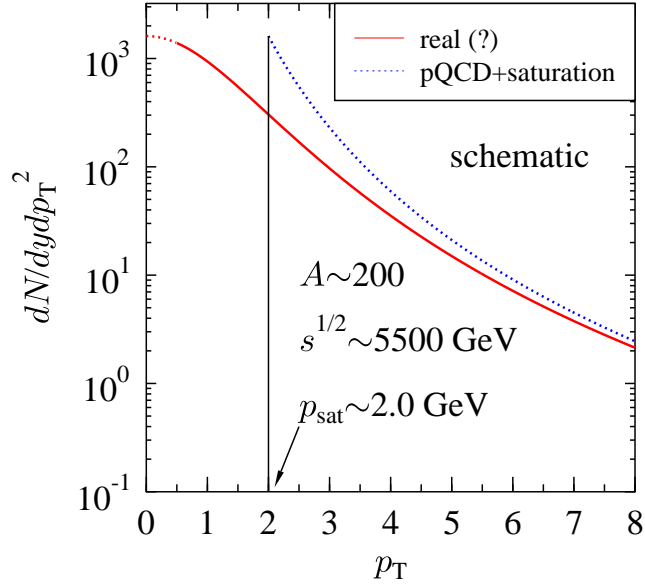}}
\vspace{-8.9cm}
\caption[a]{\protect \small The qualitative
$p_T$ distribution of gluons
produced in a heavy ion collisions at $\roots=5500$ GeV (solid curve).
Only the integral over $p_T$ has physical meaning; the
saturation+pQCD model produces the same integral by integrating
over the dashed curve computed using pQCD but
with the lower limit determined by
the saturation condition \nr{satcond}.
}
\la{mantra}
\end{figure}

To approximately solve Eq.~\nr{satcond} for $p_0=\psat$
one may neglect the $p_0$-dependence of $\sigma$ 
and approximate $\alpha_s^2(p_0)\sim
1/p_0^{2\xi},\,\,\xi\approx0.44$. This numerical value actually
contains also a contribution from the $p_T$-integration
\cite{ekt2}. Then,
\ba
\psat(y)&\sim&A^{1/(6+3\xi)}\roots^{\,\delta/(2+\xi)}
e^{-y^2/(4(2+\xi)\sigma^2)}\la{psaty}\\
&\sim&0.208\,\gev A^{0.128}\roots^{\,0.191}
e^{-y^2/(4(2+\xi)\sigma^2)},\la{psatynum}\\
dN/dy&\sim&A^{(6+2\xi)/(6+3\xi)}\roots^{\,2\delta/(2+\xi)}
e^{-y^2/(2(2+\xi)\sigma^2)}\\
&\sim&1.383 A^{0.922}\roots^{\,0.383}
e^{-y^2/(2(2+\xi)\sigma^2)},\la{ny}
\ea
where the numerical values computed in \cite{ekrt} with 
shadowing are also given.

The scaling exponents of $A$ and $\roots$ are the ones from
\cite{ekrt,ekt2}; the rapidity dependence is new.  Note how the
saturation condition leads to $N\sim A^a,\, a\approx1$ \cite{bm},
though in the independent collision limit $N\sim A^{4/3}$. Similarly,
the saturation condition has a significant effect on the rapidity
distribution of saturated particle production: it is wider than that
of elementary subprocesses by a factor $\sqrt{2+\xi}\approx 1.6$.

In the numerical computation, one first evaluates
$dN/dy(p_0)=T_{AA}(0)d\sigma_\pqcd/dy(p_0)$, The result is shown in
Fig.~\ref{viuhka} together with a gaussian fit
$\sim\exp(-y^2/(2\sigma^2))$. The distributions get narrower when
$p_0$ increases: at $\roots=5500$ GeV $\sigma(p_0)\approx6.6-p_0$.

\begin{figure}[hbt]
\vspace{-1cm}
\hspace{0cm}
\centerline{\hspace{0.5cm}\epsfysize=9.5cm
\epsffile{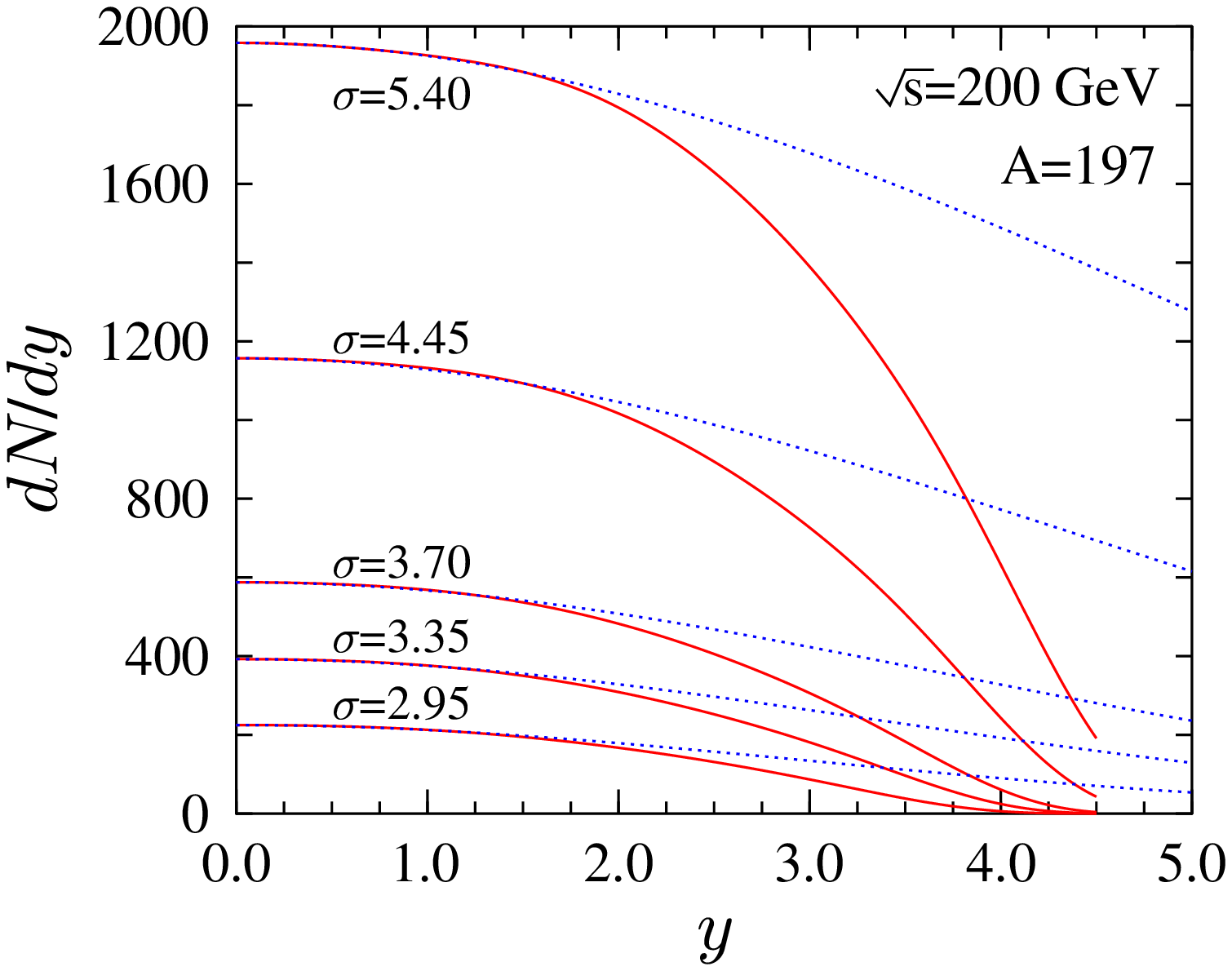}
\hspace{-1cm}\epsfysize=9.5cm
\epsffile{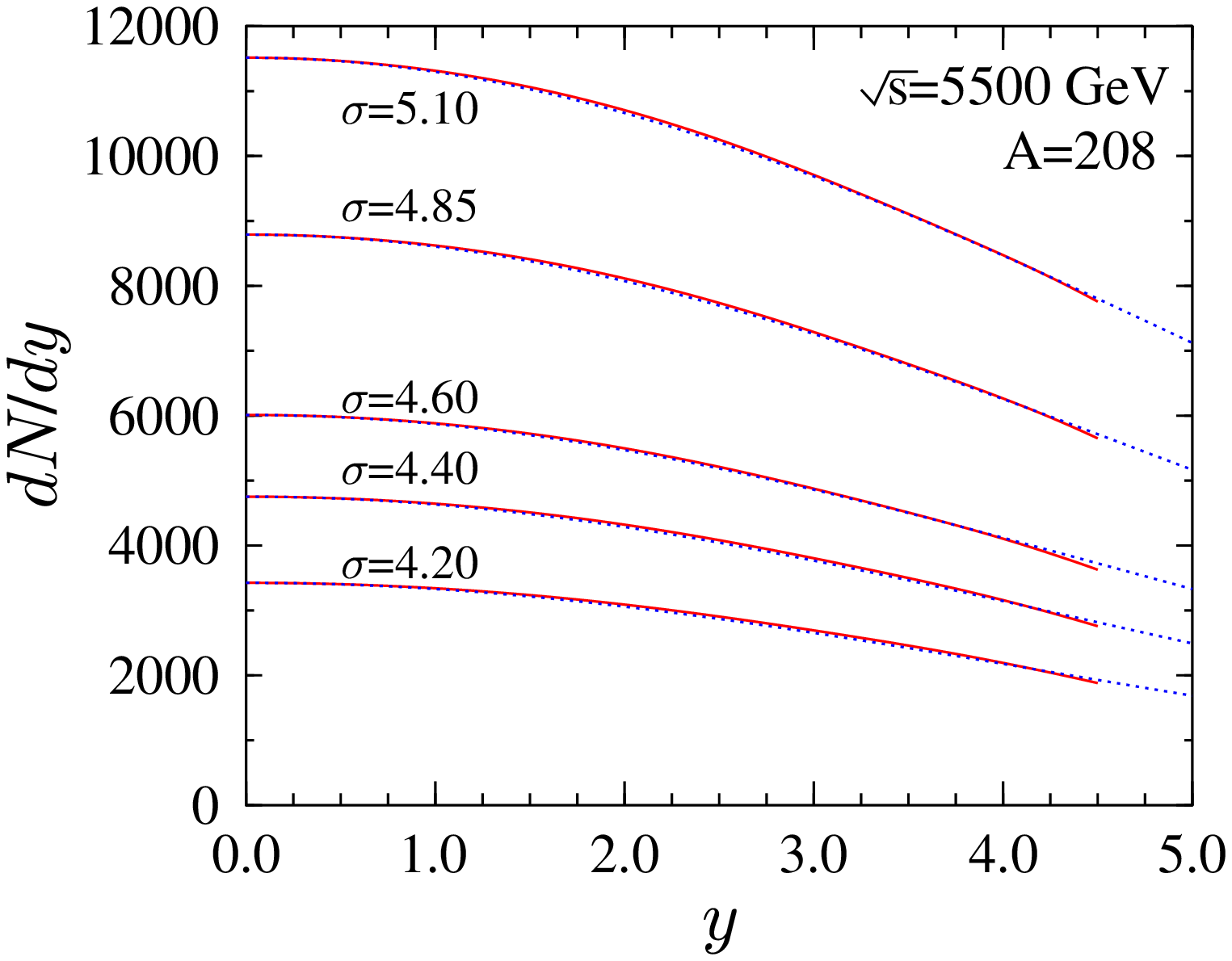} }
\vspace{-2cm}
\caption[a]{\small Plots of $dN/dy=T_{AA}(0)d\sigma_\pqcd/dy(p_0)$ for $\roots=
200$ GeV (from top to bottom $p_0=1.0,1.2,1.5,1.7,2.0$ GeV)
and 5500 GeV (from top to bottom $p_0=1.5,1.7,2.0,2.2,2.5$ GeV).
The dashed lines show a gaussian fit $\sim\exp(-y^2/(2\sigma^2))$.
 } \la{viuhka}
\end{figure}

Using the numerically computed $dN/dy(p_0)$ the saturation condition
operates as shown in Fig.~\ref{satu} and leads to $N_{AA}(y)$ also in
Fig.~\ref{satu}. In agreement with the analytic argument in
Eq.~\nr{ny} a very broad distribution near $y=0$ is obtained:
\ba
\sigma&=& 5.8\qquad \roots=130\,\,\,\gev,\\
&=& 5.9\qquad \roots=200\,\,\,\gev,\\
&=& 6.8\qquad \roots=5500\,\gev.
\ea

\begin{figure}[hbt]
\vspace{-1cm}
\hspace{0cm}
\centerline{\hspace{0.5cm}\epsfysize=9.5cm
\epsffile{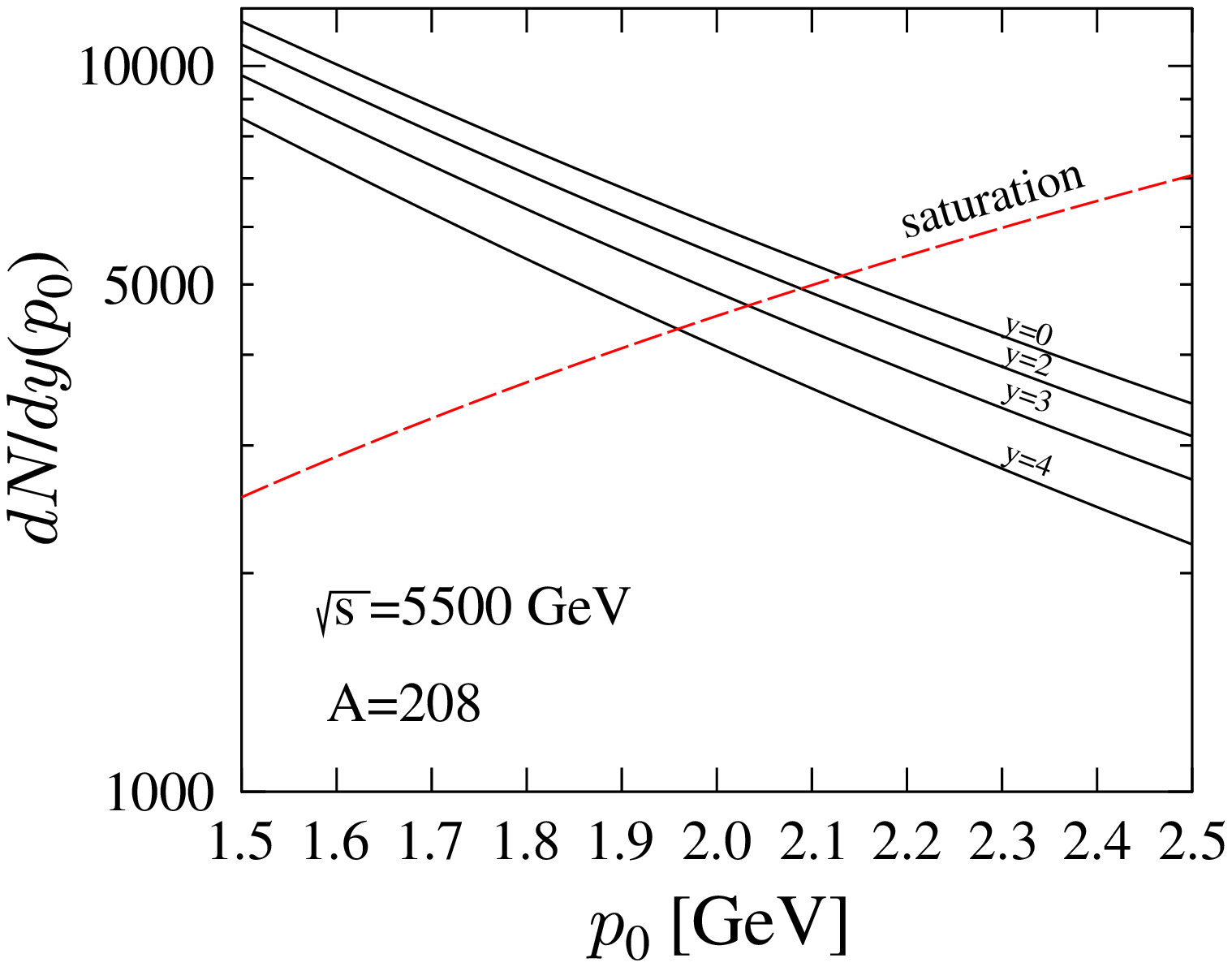}
\hspace{-1cm}\epsfysize=9.5cm
\epsffile{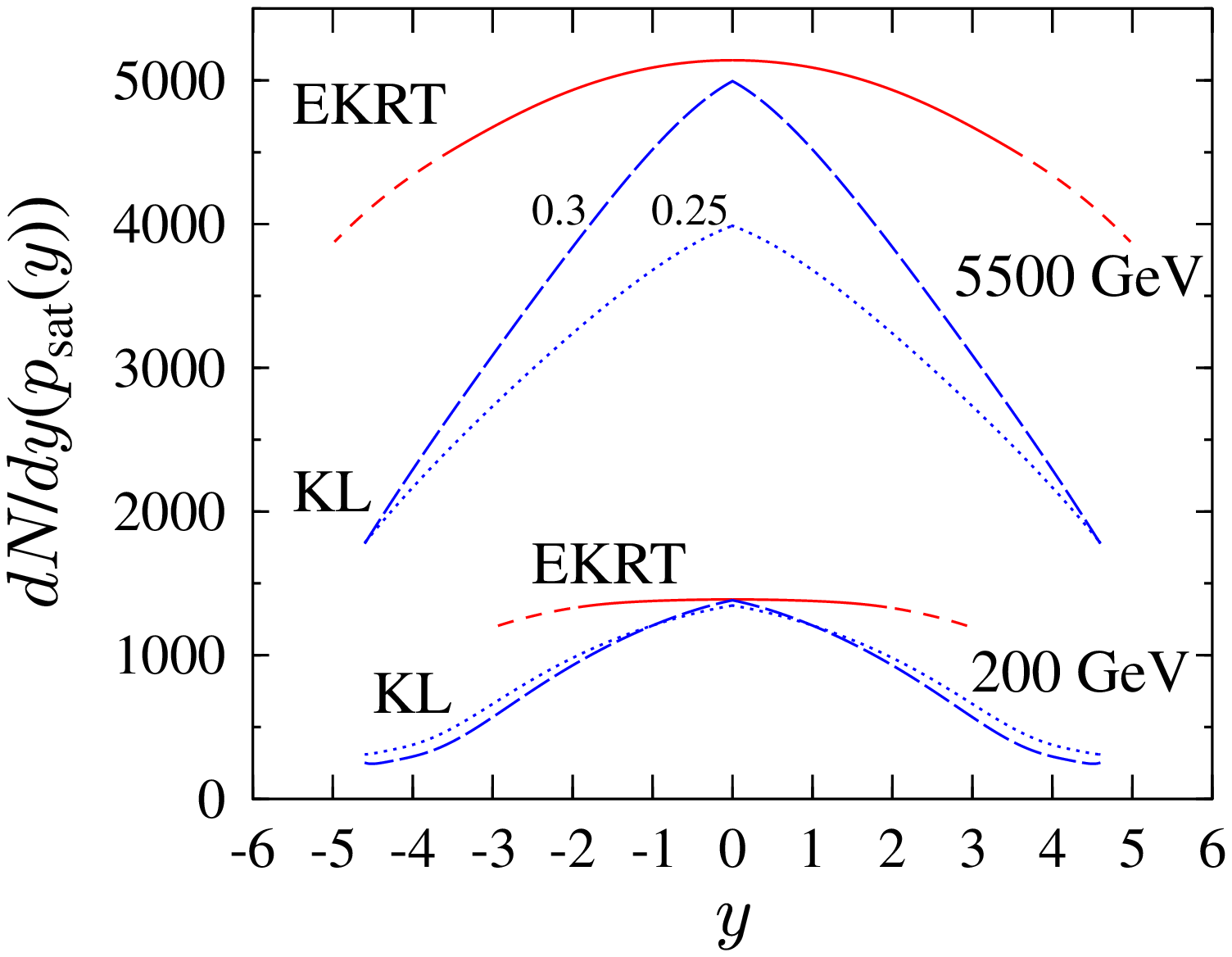} }
\vspace{-2cm}
\caption[a]{\small The left panel shows how the saturation condition
\nr{satcond} operates: the dashed line shows its RHS $p_0^2R_A^2$, the
solid lines show the LHS $T_{AA}(0)d\sigma_\pqcd(p_0,y)/dy$ at
$\roots=5500$~GeV for various $y$ as a function of $p_0$. The point of
intersection gives both the value of $\psat(y)$ and of
$dN/dy(\psat(y))$, the value of which is plotted in panel at right for
central Pb-Pb collisions at $\sqrt s = 5500$ GeV and Au-Au at 200 GeV.
The dotted and dashed curves show the prediction of \cite{kl} for
$\lambda=0.25,\,0.3$ respectively.
 } \la{satu}
\end{figure}

The form of the rapidity dependence of the saturation scale
proposed in \cite{kl,kln} is markedly different. One first notes 
that the conjectured
saturation scale as determined from deep inelastic scattering
\cite{golec-biernat} can be parametrised as
\be
Q_s^2(x)=Q_0^2\left({x_0\over x}\right)^\lambda,
\la{Qskl}
\ee
where $Q_0\approx$ 1 GeV, $x_0\approx3\cdot10^{-4}$ and
$\lambda\approx0.3$.  To relate the DIS Bjorken variable $x$ to the
A+A gluon production variables, one further writes that $x\sim
e^{y-y_B}$. Then $Q_s^2(y)\sim\roots^\lambda e^{-\lambda
|y|}$. Including only the region in which both initial gluons in the
2$\to$1 process are saturated, one obtains
\be
{dN\over dy}\sim Q_s^2(y) \sim\roots^\lambda e^{-\lambda |y|}.
\label{2to1pred}
\ee 
Thus the rapidity distribution will have a sharp peak at $y=0$, as
shown in the right panel of Fig.~\ref{satu}.  In Ref. \cite{kl} it is
assummed that the pseudorapidity distribution of produced gluons
(virtual, $m_g^2\approx Q_s\cdot 1$ GeV, $p_T\approx Q_s$) directly
give the pseudorapidity distributions of pions in the final state,
apart from a multiplicative normalization factor $c$.  Thus $dN_{\rm
ch}/d\eta\approx c\cdot dN_g/d\eta$, where $c$ contains factors from
the actual gluon production, such as the gluon liberation constant
$c_g={\cal O}(1)$ and possible factors related to the normalization
and approximations made for the unintegrated gluon densities. The
constant $c$ also includes factors related to hadronization, decays
and particle content in the final state. Such factors are $c_{\pi/g}$
which indicates how many pions come from each virtual gluon, and a
factor $c_{\rm ch}$ accounting for the conversion of total to charged
multiplicity. It should be noted that when plotting the rapidity
distributions of \cite{kl} for the produced gluons in Figs.~3 (right
panel) and in Fig.~5 (left panel), we have fixed the normalization
constant $c=c_{\rm ch}\cdot c_{\pi/g}\cdot c_g$ based on the measured
$dN_{\rm ch}/d\eta$ \cite{phobos4}, and used $c_{\rm ch}\approx 2/3$.
As our primary goal here is to compare the shapes of the rapidity
spectra of the two approaches, we have not tried to undo the effect of
the factor $c_{\pi/g}$. As the virtuality of the produced gluons in
\cite{kl} is taken to be larger than $m_{\rho}$, we expect that the
actual number of initially produced gluons in the saturation model of
\cite{kl} can be down by $c_{\pi/g}\sim2$ relative to what is shown in
Figs.~3 and 5.

One may observe that in the final state saturation calculation one
does not at all need the distribution functions within the saturation
region as defined by Eq.~\nr{Qskl}.  To check the effect of saturated
distribution functions on the 2$\to$2 model one can perform the
following computation.  Including only the dominant gluons one has
\be
{d\sigma\over dy d^2p_T}\approx
K\int dy_2\,x_1g(x_1,p_T^2)\,x_2g(x_2,p_T^2)
{9\alpha_s^2\over2p_T^4},
\la{incl}
\ee
where $x_1=p_T/\roots\cdot(e^y+e^{y_2}),\,\,
x_2=p_T/\roots\cdot(e^{-y}+e^{-y_2})$. A simple model for
saturated distribution functions (see \cite{kl}) would be
\be
xg(x,p_T^2)=C{S_A\over\alpha_s}\left\{ p_T^2\cdot 
\Theta\left[Q_0^2({x_0\over x})^\lambda -p_T^2\right]+
Q_0^2({x_0\over x})^\lambda 
\Theta\left[p_T^2-Q_0^2({x_0\over x})^\lambda\right]\right\}
\ee
with $S_A\approx\pi R_A^2$, $C$ = constant, and where implicitly
$Q_0^2\sim A^{1/3}$. Integration over $y_2$ and $p_T$ then gives a
$y$-distribution which is broad and very similar to those in
Figs.~\ref{viuhka} and \ref{satu}. We thus conclude that the broad
$y$-distribution is not due to the use of distributions in the DGLAP
region, it is a property of the 2$\to$2 model as such.

The $\roots$ and $y$ dependences of the result (\ref{2to1pred}) are
sensitive to the value of the parameter $\lambda=0.25\dots 0.30$,
especially when one approaches LHC energies.  Furthermore this
$\lambda$ interval is based on the analysis in
Ref. \cite{golec-biernat} which neglects the scale evolution of the
gluon distribution in the non-saturated region on which the saturated
gluon distribution has to be matched in the vicinity of the 'critical
line' $Q_s^2(x)$ defining the transition region from saturated to
non-saturated gluon densities.  An interesting recent analysis in
\cite{stasto} suggests that one might be led to consider values as
large as $\lambda\sim 0.5$. This is similar to the behaviour of the
gluon distribution at the final state saturation scale \cite{ekt2}.

\section{Energy conservation}
The results \nr{psaty}-\nr{ny} lead to a rather striking powerlike
growth with energy, especially for the transverse energy
$\sim\psat(y)N(y)$. Let us thus estimate when the total amount
of energy within some rapidity range $-y_\rmi{up}<y<y_\rmi{up}$ is
less than some fraction, 40\%, say, of
the total energy $A\roots\approx Ae^{y_\rmi{B}}$
available (GeV units are used), neglecting first the weak
rapidity dependence.
A simple analytic estimate of this is
($c_N,c_p$ are the constants in \nr{psatynum},\nr{ny}):
\ba
E(|y|\le y_\rmi{up})&\approx&\int_{-y_\rmi{up}}^{y_\rmi{up}}dy
\frac{dN}{dy}\bigg|_{y=0}\psat(y=0)\cosh y
\approx \frac{dN}{dy}\bigg|_{y=0}\psat(0) e^{y_\rmi{up}}\\\nonumber
&\approx&
c_Nc_p A^{(7+2\xi)/(6+3\xi)}\roots^{\,3\delta/(2+\xi)}e^{y_\rmi{up}}
<0.4A\roots
\ea
giving
\be
y_\rmi{up}+\log(c_Nc_p A^{(1-\xi)/(6+3\xi)}) < 
y_B(1-{3\delta\over 2+\xi})+\log0.4.
\la{ymax}
\ee
The two logs are constants of order one, 
but the qualitatively important factor
is $y_B(1-3\delta/(2+\xi))\approx y_B(1-0.62)$; it expresses the
fact that with increasing beam energy particle production in
the saturation + pQCD model will exhaust the total available energy
within a fixed fraction
of the beam rapidity $y_B$. Beyond that rapidity this independent
subcollision model cannot be valid and
the physical mechanism has to change completely.
The same effect is shown more quantitatively in Fig.~\ref{econsv},
where we have computed the total produced energy  as  
\be
E(y_{\rm up})=\int_{-y_{\rm up}}^{y_{\rm up}} dy\int_{\psat(y)} dp_T 
 \frac{dN}{dp_Tdy}\cdot p_T\cosh y,
\la{Etot}
\ee
The rapidity distributions of produced partons are obtained from 
Eq.~(\ref{minijets}), as discussed in detail in Ref. \cite{EKbaryons}.
Keeping the saturation scale fixed at $\psat(y=0)$ gives the dotted
lines, not qualitatively different from the solid lines obtained with
the rapidity dependent saturation.

\begin{figure}[hbt]
\vspace{4cm}
\centerline{ \epsfysize=8cm \epsffile{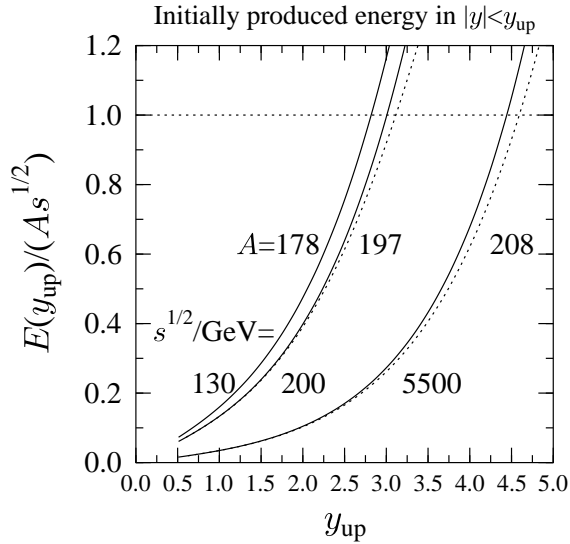}\hspace{7cm}}
\vspace{-4cm}
\caption[a]{\small The total energy produced in the saturation + pQCD
model up to $|y|<y_\rmi{up}$ relative to the cms-energy in central
$AA$ collisions. The cms-energies and nuclei are denoted in the
figure. ``Initial'' refers to time after formation before any
collective expansion. The solid curves are computed from Eq.~(\ref{Etot}) 
with $\psat(y)$, and the dotted curves with a fixed $\psat(y=0)$
as the lower limit of $p_T$-integration.  }
\la{econsv}
\end{figure}

The saturation+pQCD model thus predicts that
the $y$-distribution $N(y)$ is very flat, gaussianlike around
$y=0$, with height increasing as
given by Eq.~\nr{ny} but extending only up to some maximum
value of $y$ imposed by energy conservation. 
We emphasize that
the saturation condition itself leads to a broadening of the
rapidity distribution.
Allowing 40\% of the total energy leads
to $|y|\lsim2$ at RHIC (with $y_B=4.93$ (5.36) for $\sqrt s=$ 130 (200) GeV) 
and $|y|\lsim 3.5$ at LHC (with $y_B=8.67$).
Beyond these values correlations between subcollisions must 
start to play an increasingly important role.

\section{Hydrodynamic evolution}
The previous considerations apply at the initial time $\tau_i(y)
\approx 1/\psat(y)$. Within the domain of validity, $|y|<y_\rmi{up}$,
there is little variation in $\tau_i(y)$. Hydrodynamic evolution in
the case of $y$-dependent initial conditions of the type obtained here
was numerically studied in \cite{ekr}. There is always significant
flow of $E_T$ from small to large rapidities, due to $pdV$ work
\cite{ekrt,errt}, but in this $y$-dependent case there is also some
flow of entropy from small to large rapidities. Typically, the entropy
density at $y=0$ decreases by about 10\%, as shown in \cite{ekr}.

We thus conclude that the time evolution of $y$-distribution is one 
more detail to be
added to the list of factors affecting the relation between initial
and final multiplicity \cite{errt}: initial and final particle-to-entropy
ratios, charged-to-neutral ratio, transverse expansion, equation of
state, decoupling effects and resonance decays. However, 
in the central region the hydrodynamic evolution affects
only the overall magnitude of the rapidity distribution
while the effect on the shape is small.

\section{Comparison with RHIC data}
We note first that the
RHIC data \cite{phobos4,brahms} gives the pseudorapidity distribution
$dN_{\rm ch}/d\eta$ while our computations so far are for $dN/dy$ at the moment
of formation of the partonic system. The conversion of 
$N$ to the observed $N_\rmi{ch}$ is simple (see end of previous
section), but the relation between $y$- and $\eta$-distributions
is more subtle \cite{errt}.

The rapidity distribution of particles is the sum
\begin{equation}
\frac{dN}{dy}\equiv \int dp_T \sum_i\frac{dN_i}{dp_T dy},
\label{dNdy}
\end{equation}
where the index $i$ runs through all particles included in the
spectrum, similarly for pseudorapidity.
  We have shown
explicitly the transverse momentum integration to emphasize that the
connection between the rapidity and pseudorapidity depends on the mass
and the transverse momentum of particles:  $y={\rm
arsinh}((p_T/m_{Ti})\sinh\eta)$.  Similarly the Jacobian
\begin{equation}
J_i(\eta,p_T,m_i)=\frac{\partial y}{\partial\eta} = \frac{p}{E_i}
\end{equation}
in the transformation between the spectra
\begin{equation}
\frac{dN}{d\eta}\equiv\int dp_T \frac{dN}{dp_T d\eta }
= \int dp_T  \sum_i J_i(\eta,p_T,m_i) \frac{dN_i(p_T,y)}{dp_T dy},
\label{ytueta}
\end{equation}
depends on the mass and the transverse momentum.
In the following we will assume that pions dominate
and consider equations for single particle only.

The pseudorapidity data \cite{phobos4} is characterized by a dip in a
central plateau of total extent of 3...4 in $\eta$. Starting from a
flat rapidity distribution, this dip results from the Jacobian in the
transformation to pseudorapidity. 
The depth of the dip is quite sensitive to particle masses and 
shapes of the transverse momentum and rapidity distributions,
variations by a factor two in the depth of the dip can easily take
place.
We also emphasize the difference between performing the
transformation using Eq.~(\ref{ytueta}) with $p_T$ and
$y$ distributions or simply approximating the Jacobian by  
$dN/d\eta=J(\eta,\langle p_T\rangle,m)dN/dy$.

\begin{figure}[hbt]
\vspace{1cm}
\centerline{ \hspace{-2cm}
\epsfysize=9cm \epsffile{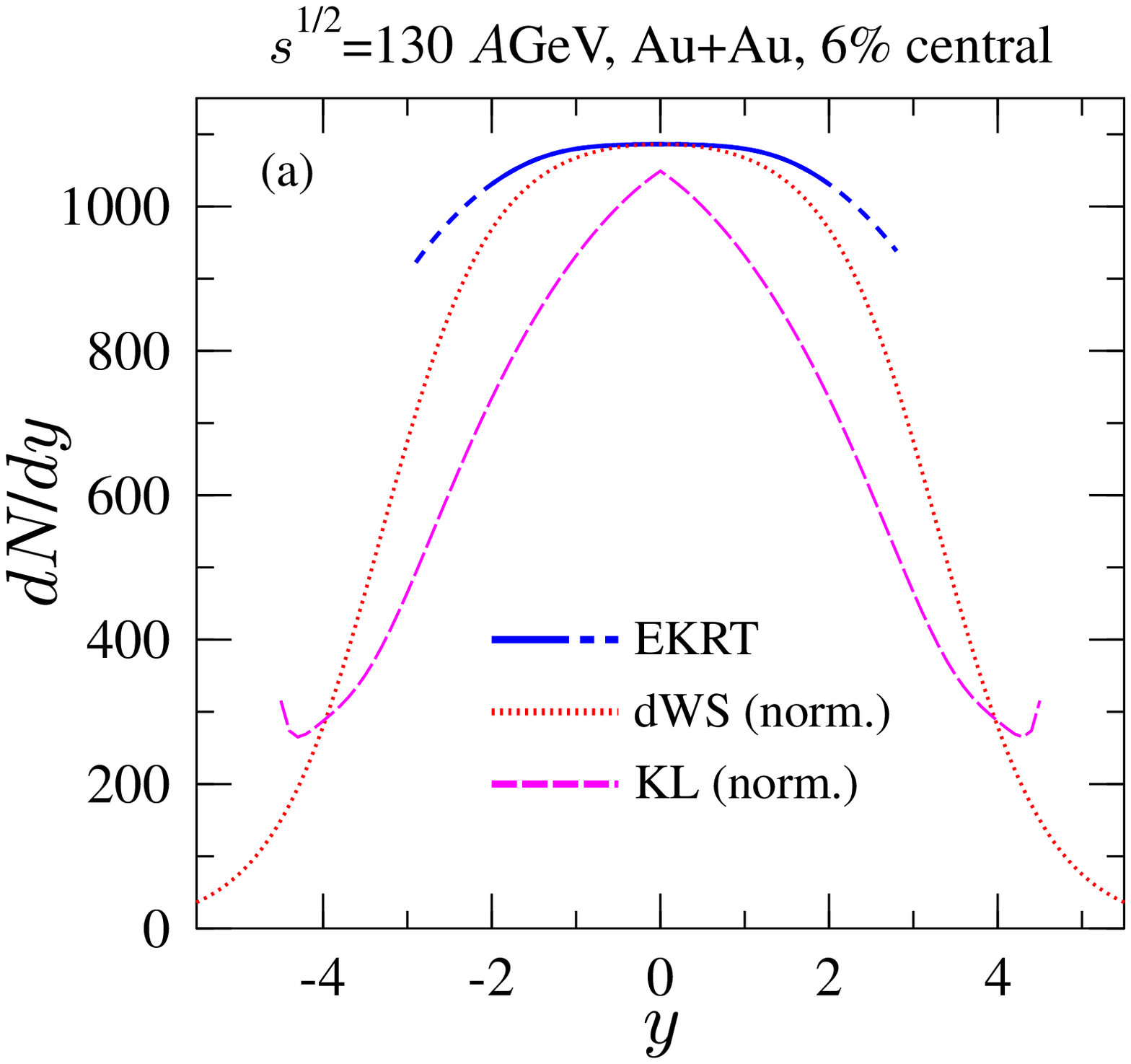} \hspace{-0.5cm}
\epsfysize=9cm \epsffile{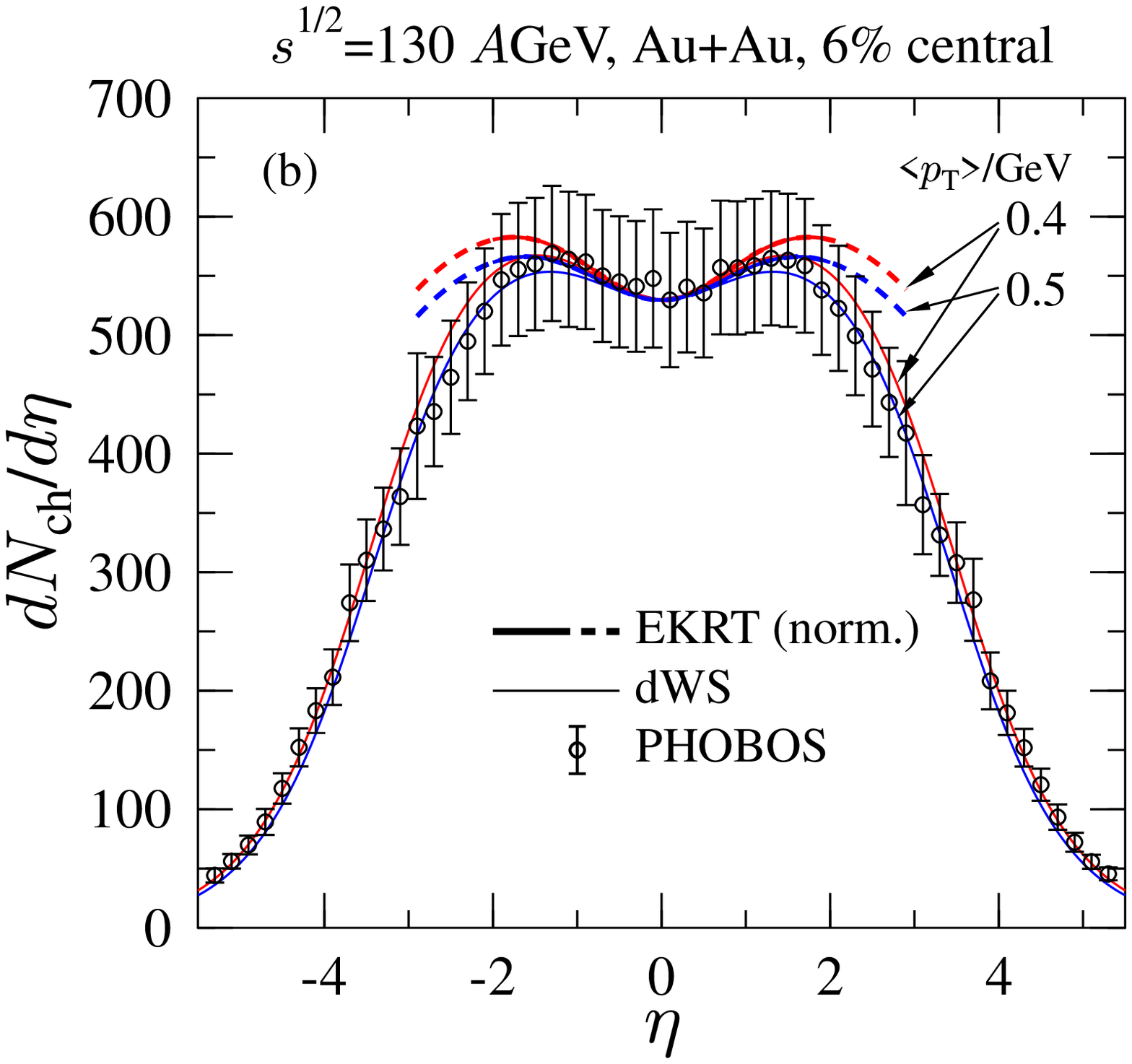}
}
\vspace{-2cm}
\caption[a]{\small (Left) Initial rapidity distributions
$dN/dy$ for the saturation+pQCD model
(solid curve, EKRT), for the double-Woods-Saxon (dWS)
parametrization \nr{FermiDirac} (dotted curve) and for the
prediction of \cite{kl} (dashed curve) (see text).
(Right) The final state pseudorapidity distributions
$dN/d\eta$ (6\% central): data \cite{phobos1}, the
saturation + pQCD model (thick solid lines, EKRT)
and the parametrisation \nr{FermiDirac} (see text).
} 
\la{dndeta130}
\end{figure}

In Fig.~\ref{dndeta130}a the results from calculations of the 
initial $dN/dy$  are shown for nearly central Au+Au collisions 
at $\roots=130$ GeV. The solid curve (EKRT) is the prediction with
$y$-dependent saturation of produced
partons, computed as in Figs.~\ref{satu}. A 6\% centrality cut
corresponds to an effective nucleus $A=178$, as discussed in
\cite{errt}.  Within the dashed part of the curve more than 40\% of
total energy has been consumed (Fig.~\ref{econsv}); the curves stop
when all the energy is consumed.  The dotted curve shows a
double-Woods-Saxon (dWS) parametrization
\begin{equation}
   {dN\over dy} =
N(0)\frac{(1+e^{-y_0/d})^2}{(1+e^{(-y-y_0)/d})(1+e^{(y-y_0)/d})}\,,
\label{FermiDirac}
\end{equation}
normalised to the EKRT result at $y=0$.  Here $y_0=3.3$ gives the
width of the distribution and $d=0.65$ the steepness at this
cutoff. This double Woods-Saxon form for $dN/dy$ is an arbitrary
parametrization designed to reproduce the pseudorapidity data
\cite{phobos4} but it has the two elements we want to emphasize, the
flatness of $dN/dy$ at small rapidities and the rapid decrease at
large values of $y$ required by energy conservation. Finally, the
dashed curve is the initially produced rapidity distribution from
\cite{kl}. The normalization of this curve was discussed in the end of
Sec. 2.

Fig.~\ref{dndeta130}b shows the data \cite{phobos4} 
with $dN_{\rm ch}/d\eta$ curves calculated from
$dN/dy$ assuming that all particles are pions
with an exponential transverse momentum distribution, 
$dN/dp_T^2dy\sim dN/dy \cdot \exp(-2p_T/\langle p_T\rangle)$,
with $\langle p_T\rangle=0.4$ GeV or 0.5 GeV. 
The thick solid lines (EKRT) are the saturation model
prediction, computed using Eq.~\nr{ytueta}. The dashed parts of the
lines correspond to those in the left panel and indicate
again where energy conservation is expected to suppress the
distributions.  The thin solid lines (dWS) are from the parametrization
\nr{FermiDirac} with $y_0=3.3$ and $d=0.65$, which reproduce the
PHOBOS data.  
In the framework of \cite{ekrt}, the initial and final state rapidity 
distributions are connected through entropy conservation by 
$dN_{ch}/dy=2/3\cdot 3.6/4.0\cdot dN/dy(\psat(y))$.
Here, in order to study the shapes of the obtained spectra, 
all the curves are normalized to the same point at $\eta=0$. 
Compared with the normalization in \cite{ekrt}, an additional 
factor 0.95 (0.92) for  $\langle p_T\rangle=0.4$ (0.5) GeV
is applied to the EKRT curves in Fig.~\ref{dndeta130}b.
The relation between the initial and final multiplicities
is studied in more detail in  \cite{errt}.

We do not show here the pseudorapidity distribution from the dashed
curve in the left panel; in \cite{kl} the $y\to\eta$ transformation
was carried out simply using a Jacobian $J(\eta,p_T,m_g)$, calculated
for virtual gluons with $m_g^2\approx Q_s\cdot 1$ GeV and $p_T\approx
Q_s$, as a multiplicative factor between $dN/dy$ and $dN/d\eta$.  With
$Q_s(y=0)=\sqrt 2$ GeV \cite{KN} the suppression from the $y$- to
$\eta$- distribution at $y=\eta=0$ is $p_T/m_T\approx 0.76$, which
turns the peak in the $y$ distribution into a dip in the
$\eta$-distribution.  The results of \cite{kl} are thus very close to
the data in Fig.~\ref{dndeta130}b.

\section{Conclusions}

We have discussed the rapidity distribution $dN/dy$ of particles
produced in very high energy central $A+A$ collisions in the
saturation+pQCD model. The distribution around $y=0$ will be a broad
gaussian with a calculable width and with the value $dN/dy(y=0)$
increasing rapidly, powerlike. However, the model can be applied only
until some value $y_\rmi{up}$ ($\approx$ 1...2 at RHIC, 3...4 at LHC),
beyond which some new type of fragmentation region dynamics, taking
into account correlations between subcollisions, must enter.  At
asymptotic energies the saturation region is parametrically favoured
by powers of $1/\alpha_s$ and this will damp the distribution at large
$y$.

The overall magnitude of $dN/d\eta$ at $\eta=0$ predicted by the
saturation + pQCD model agreed very well with RHIC experiments for
central collisions. The centrality dependence was also reproduced on
the 10\% level within $N_{\rm part}\lsim100$.  A more reliable
calculation on the $y$-dependence of initial gluon production can be
made only after a better understanding of the correlations among
subprocesses in the fragmentation and near fragmentation regions has
been achieved.  At present one may note that within the domain of
validity of the model, before energy conservation makes the assumption
of independent subcollisions invalid, there is agreement with
data. The width of the relevant $y$-range at RHIC is narrow, but
becomes appreciable at LHC energies. It will be very interesting to
see how a rapid increase at $y=0$ and dynamics in the fragmentation
regions will be related.

\vspace{1cm}
{\bf Acknowledgements}
We thank M. Gyulassy for suggesting this computation many years
ago, and D. Kharzeev  and N. Armesto for discussions.
Financial support from the Academy of Finland
(grants No. 163065, 77744 and 50338) is gratefully acknowledged.

\end{document}